\documentclass[pra,aps,twocolumn,groupedaddress]{revtex4}
\usepackage{amssymb,amsmath,amsfonts,epsfig,graphicx,latexsym,bm,color}

\newcommand{\beq}{\begin{eqnarray}}
\newcommand{\eeq}{\end{eqnarray}}

\begin{document}
\title{Non-destructive monitoring of Bloch oscillations in an optical cavity}
\author{H. Ke{\ss}ler$^1$, J. Klinder$^{1}$, B. P. Venkatesh$^2$, Ch. Georges$^1$, and A. Hemmerich$^{1,3}$}
\affiliation{$^1$ Institut f\"{u}r Laser-Physik, Universit\"{a}t Hamburg, 22761 Hamburg, Germany \\
$^2$Institute for Theoretical Physics, University of Innsbruck, A-6020 Innsbruck, Austria \\
$^3$Wilczek Quantum Center, Zhejiang University of Technology, Hangzhou 310023, China}

\begin{abstract}
Bloch oscillations are a hallmark of coherent wave dynamics in periodic potentials. They occur as the response of quantum mechanical particles in a lattice if a weak force is applied. In optical lattices with their perfect periodic structure they can be readily observed and employed as a quantum mechanical force sensor, for example, for precise measurements of the gravitational acceleration. However, the destructive character of the measurement process in previous experimental implementations poses serious limitations for the precision of such measurements. In this article we show that the use of an optical cavity operating in the regime of strong cooperative coupling allows one to directly monitor Bloch oscillations of a cloud of cold atoms in the light leaking out of the cavity. Hence, with a single atomic sample the Bloch oscillation dynamics can be mapped out, while in previous experiments, each data point required the preparation of a new atom cloud. The use of a cavity-based monitor should greatly improve the precision of Bloch oscillation measurements for metrological purposes.
\end{abstract}

\maketitle

As originally pointed out by Bloch and Zener \cite{Blo:29, Zen:34}, a coherent wave packet in a lattice responds to the presence of a force by an oscillatory rather than a uniform motion with the fundamental frequency $\Omega_B \equiv F d / \hbar$ exclusively depending on the size of the force $F$ and the lattice constant $d$. Such Bloch oscillations are a paradigm of coherent wave dynamics in periodic potentials. While in electronic lattice gases Bloch oscillations are covered by incoherent scattering, neutral atoms confined in optical lattices have proven nearly ideal to study this quantum phenomenon and to use it as a precision force sensor, for example for measurements of the gravitational acceleration \cite{Ben:96, Mor:01, Roa:04, Fer:06, Gus:08, Hal:10, Pol:11}. The conventional method to map out the Bloch frequency $\Omega_B$ in an optical lattice requires many successive destructive position measurements of the oscillating atomic wave packet, each of which requires the production of a new atomic sample. This inefficient data acquisition method seriously limits the attainable precision \cite{Fer:06, Pol:11,Tar:12}. 

\begin{figure}
\includegraphics[scale=0.32, angle=0, origin=c]{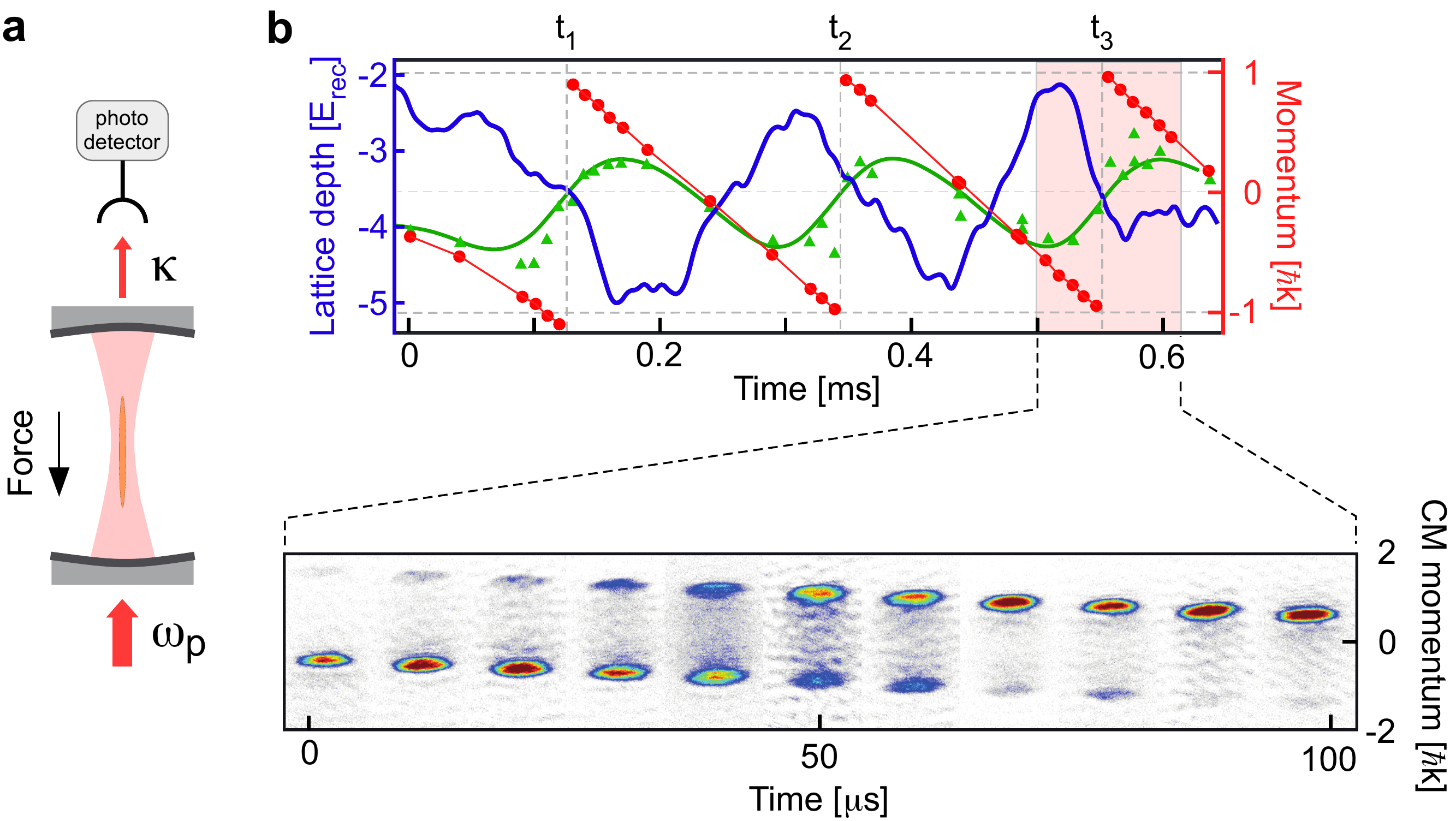}
\caption{(a) Sketch (not drawn to scale) of a BEC placed inside a standing wave cavity, which is axially pumped at frequency $\omega_p$. A magnetic force is applied along the cavity axis. (b) Typical data set: the blue thick line shows the intra-cavity intensity (parametrized in terms of the associated (negative) intra-cavity light-shift), observed during ca. $600\,\mu$s by the photo detector in (a) for the same BEC in single experimental run. The red disks (connected by solid red straight line segments for eye guiding) show the positions of the density maxima of the atomic cloud versus the holding time in the lattices. The green triangles show the corresponding centre-of-mass of the atomic cloud. The green solid lines shows a calculation for $\delta_{c} = -5.8 \, \kappa$. Each data point requires preparation of a new BEC. For the section of the time-axis highlighted by a reddish background images of the BECs after ballistic expansion are shown associated with the data points.}
\label{fig:Fig.1}
\end{figure}

Recent theoretical work has proposed to form the optical lattice in an externally pumped Fabry-P{\'e}rot cavity \cite{Ven:09, Ven:13} or in a ring resonator \cite{Ped:09, Gol:14, Sam:14} and to use the light leaking out through one of the cavity mirrors as a non-destructive monitor for the oscillating atoms. The appearance of a notable signature in the intensity (thus avoiding the extra expense of heterodyne techniques for phase detection) relies on sufficient back-action of the atoms upon the intra-cavity light field. This requires operation of the cavity in the regime of strong cooperative coupling, which has been experimentally demonstrated in standing wave \cite{Wol:12, Kes:14} as well as in ring resonators \cite{Nag:03, Els:04}. In a simplified picture of conventional Bloch oscillation dynamics the atomic wave packet may be approximated by a Bloch function. The force leads to a linear increase of the quasi-momentum. When the edge of the first Brillouin zone (FBZ) is reached, Bragg scattering becomes resonant and the wave packet is reflected. At zero quasi-momentum, the Bloch function is nearly constant with only a small modulation commensurate with the lattice potential. As the edge of the FBZ is approached and Bragg scattering sets in, the modulation depth grows, which in presence of a cavity of sufficient finesse leads to a notable shift of the cavity resonance frequency and hence to a change of the in-coupled intensity. This leads to an observable modulation of the transmitted intensity at the Bloch frequency $\Omega_B$. As pointed out in Refs. \cite{Ven:09, Ped:09}, $\Omega_B$ is not modified by the influence of the resonator such that no systematic errors are expected.

\textbf{Experiment}
The experimental set-up is sketched in Fig.~\ref{fig:Fig.1}(a). A cigar-shaped Bose-Einstein condensate (BEC) of $N_a \approx 5-10 \times 10^4$ $\mathrm{^{87}Rb}$-atoms (prepared in the upper hyperfine component of the ground state $|{F=2,m_F=2}\rangle$) with Thomas-Fermi radii $(R_x, R_y, R_z) = (3.1, 3.3, 26.8)\,\mu$m is held in a magnetic trap (trap frequencies $\omega_{\mathrm{x,y,z}} / 2\pi = (215.6, 202.2, 25.2)\,$Hz). The BEC is superimposed to a longitudinal mode ($\approx\, 32\,\mu$m waist) of a high finesse (finesse = 344.000, Purcell factor = 44 \cite{Pur:46, Kle:81}, field decay rate $\kappa = 2\pi \times 4.45\,$kHz) optical cavity extending along the $z$-axis. The cavity can be pumped axially with a laser beam well matched to a longitudinal mode. This pump beam operates at the wavelength $\lambda = 803\,$nm, i.e., at large detuning to the negative side of the principle fluorescence lines of rubidium at $780\,$nm and $795\,$nm. For a uniform atomic sample and left circularly polarized light of the pump beam, the cavity resonance frequency is dispersively shifted with respect to that of the empty cavity by an amount $\delta_{-} = \frac{1}{2} N_a \, \Delta_{-}$ with an experimentally determined light shift per photon $\Delta_{-} \approx \,-2\pi \times 0.36\,$Hz. With $N_a = 5 \times 10^4$ atoms $\delta_{-} = -2\pi\times 9$~kHz, which amounts to $-2\,\kappa$, i.e., the cavity operates in the regime of strong cooperative coupling. After BEC preparation, a typical experimental sequence comprises the following steps: 1. the intensity of the pump beam is ramped up (in $1\,$ms) to a desired value in order to initialize the lattice potential. Simultaneously, its frequency detuning with respect to the resonance of the empty cavity $\delta_{c}$ is tuned from about $- 45 \kappa$ to a desired near-resonant value. 2. The magnetic trap potential is replaced by a constant magnetic force. The switching time of the currents involved is $50\, \mu$s. 3. After a variable waiting time, the pump beam is rapidly extinguished (in $0.3\, \mu$s) such that the intra-cavity intensity exponentially decays with a time-constant of $\tau_c = 1/2\kappa = 18\, \mu$s. Simultaneously, the magnetic force is switched off in about $50\, \mu$s (current switch-off time). 4. Finally, after a ballistic time of flight of $32\,$ms the atomic cloud is imaged and the position of the density maximum of the atomic cloud is determined. During the entire protocol the intensity transmitted through one of the cavity mirrors is recorded on an avalanche photo diode.

\begin{figure}
\includegraphics[scale=0.45, angle=0, origin=c]{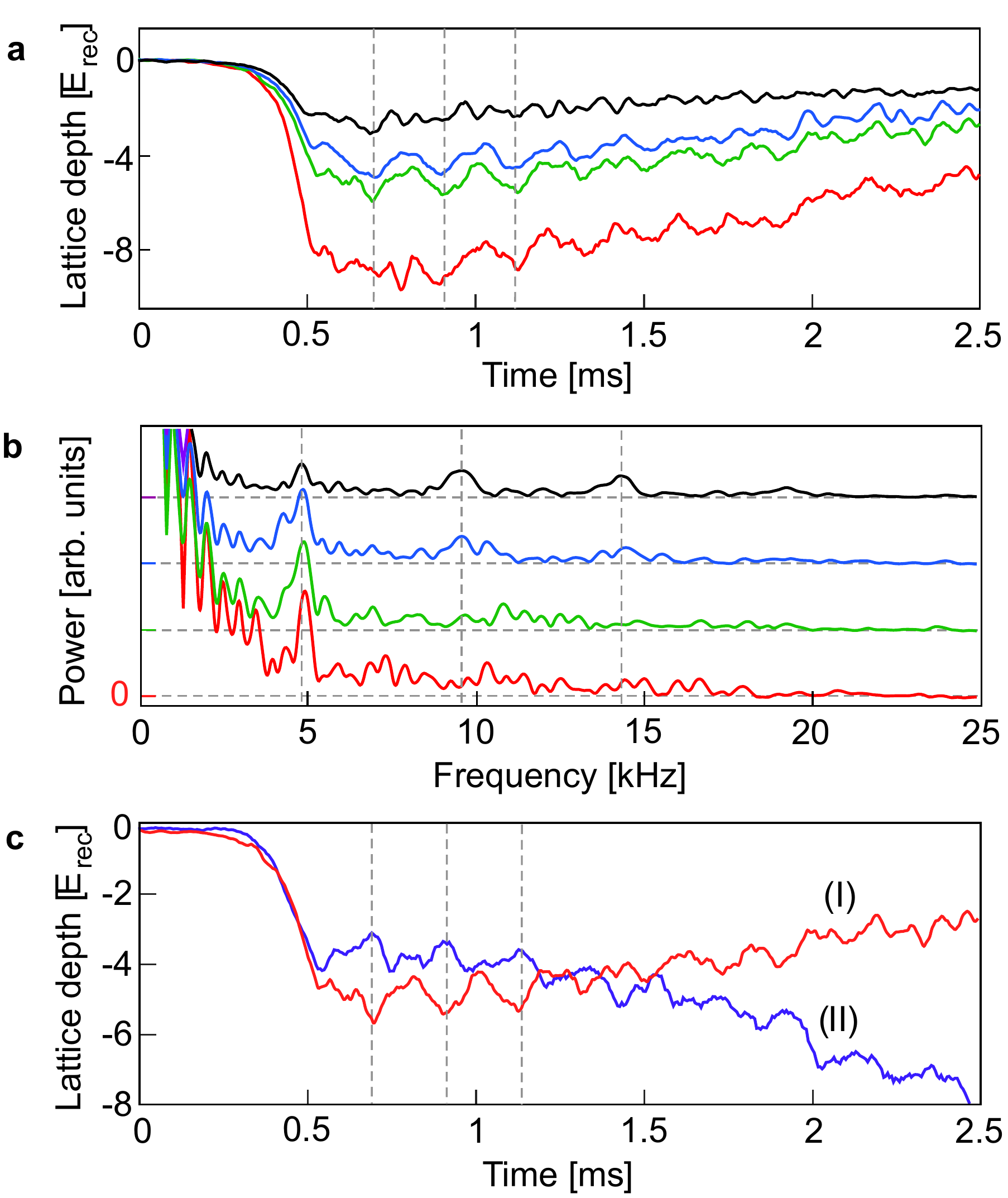}
\caption{(a) Bloch oscillations observed in the light transmitted through the cavity, given in units of the induced lattice depth. The negative pump detuning is $\delta_{c} = -7.2 \kappa$. The colored traces differ with respect to the incident pump strengths yielding different initial lattice depths $U_0$. (b) Power spectra of the graphs shown in (a). Vertical gray dashed lines mark the Bloch frequency and higher harmonics. Horizontal gray dashed lines mark the zeros of the spectra. The total particle number in (a) and (b) was about $10^5$ corresponding to $\delta_{-} = -4\,\kappa$. (c) Bloch oscillation signals for $\delta_{c} = -7.2\,\kappa$ and $N_a = 10^5$ for the red trace (I) and $\delta_{c} = -3.7 \,\kappa$ and $N_a = 8.5 \times 10^4$ for the blue trace (II). The initially prepared lattice depths $U_0$ are $-5\,E_{\mathrm{rec}}$ and $-3.5\,E_{\mathrm{rec}}$, respectively. At later times the lattice depth decreases for the former and increases for the latter case. The vertical gray dashed lines indicate the times when the BEC reaches the edge of the FBZ. Each trace in (a) and (c) represents an average over 10-20 experimental runs.}
\label{fig:Fig.2}
\end{figure}

\textbf{Observations} In Fig.~\ref{fig:Fig.1}(b) a typical data set is shown. The blue thick line shows the intra-cavity intensity, observed during ca. $600\,\mu$s by the photo detector in Fig.~\ref{fig:Fig.1}(a) for the same BEC. An oscillation at about 4.5 kHz is observed, which corresponds to the Bloch frequency adjusted via the magnetic field gradient applied in addition to the gravitational force. The red disks (connected by red solid lines for eye guiding) show the positions of the density maxima of the atomic clouds recorded after variable holding times in the lattice, shutoff of the force and the lattice potential and subsequent ballistic time-of-flight. The green triangles show the centre-of-mass of the atomic cloud associated with the red disks. The red disks and green triangles also show the Bloch oscillation period, however, in contrast to the intra-cavity intensity, here each data point requires preparation of a new BEC. The solid green line shows an associated calculation described below. For a portion of the data points highlighted by a reddish background, single-shot images of the BECs after ballistic expansion are shown in the lower half of (b). The vertical dashed gray lines mark the turning times $t_{\nu}, \nu \in \{1,2,3\}$ of the Bloch oscillation, the horizontal gray lines mark the expected edges of the FBZ. It is seen that $t_3 - t_2 < t_2 - t_1$, which we attribute to the fact that the magnetic field gradient used to generate the force needs a few hundred microseconds to completely settle. One may also notice that the first oscillation is slightly shifted with respect to the FBZ. Since sample preparation is by far the most time-consuming step in the experimental protocols of cold atom experiments, the availability of a real-time read-out of the Bloch frequency should allow to greatly speed up data acquisition in metrology experiments. In our implementation, the number of observable Bloch oscillations is severely limited by an extensive binary collision rate that leads to decoherence and heating. This can be avoided in experiments optimized for a precise determination of the Bloch frequency by adjusting zero scattering length exploiting a Feshbach resonance \cite{Koh:06, Gus:08}.

The signature of Bloch oscillations in the cavity output depends on the depth of the induced intra-cavity lattice determined by the detuning and strength of the incident pump wave. Depending on the lattice depth the observed oscillatory signal comprises different admixtures of higher harmonics of the Bloch frequency. At low lattice depths, the atomic wave packet performs a uniformly accelerated motion interrupted by Bragg reflection processes, only setting in when the edge of the FBZ is reached. Each atom in the BEC scatters one photon yielding $2 \hbar k$ momentum transfer (with $k = 2\pi / \lambda$). Hence the mean atomic velocity shows a saw-tooth like time-dependence comprising significant higher harmonic contributions. For deeper lattices, Bragg scattering can already set in, when the FBZ edge is not quite reached and the mean velocity acquires a more sinusoidal time-dependence. This is seen in Fig.~\ref{fig:Fig.2}(a) and (b), where Bloch oscillation signals observed in the light transmitted through the cavity are shown for different lattice depths and fixed negative detuning $\delta_{c} = -7.2\, \kappa$. For the lowest lattice depth (black traces) up to third order harmonic contributions can be identified in Fig.~\ref{fig:Fig.2}(b). 

The dependence of the Bloch oscillation signal upon $\delta_{c}$ is determined by whether the pump frequency $\omega_{p}$ is adjusted to the red or blue side of the effective cavity resonance frequency in presence of the BEC $\omega_{\mathrm{eff}} = \omega_{c} +  N_a \Delta_{-} \langle \cos^2(k z) \rangle_{\psi}$, where the bunching factor $\langle \cos^2(k z) \rangle_{\psi}$ measures the degree to which the BEC wave function $\psi$ forms a lattice commensurable with the cavity mode intensity pattern. In Fig.~\ref{fig:Fig.2}(c) Bloch oscillation signals for two different detunings are compared. For graph (I) (red trace) the choice of $\delta_{c} = -7.2 \kappa$ corresponds to red detuning with respect to $\omega_{\mathrm{eff}}$, i.e., $\omega_{p} < \omega_{\mathrm{eff}}$. Graph (II) with $\delta_{c} = -3.7 \kappa$ (blue trace) realizes the blue detuned case, i.e., $\omega_{p} > \omega_{\mathrm{eff}}$. The BEC starts its Bloch oscillation in the center of the FBZ near $t=0.5\,$ms, where its wave function (assumed to be approximated by the 
Bloch function at zero quasi-momentum) consists of a constant off-set and a small lattice-periodic addition. As it moves towards the edge of the FBZ, the contrast of the density modulation (i.e., the bunching factor $\langle \cos^2(k z) \rangle_{\psi}$) grows such that Bragg scattering is increasingly supported. Hence, the cooperative atom-cavity coupling increases and accordingly $\omega_{\mathrm{eff}}$ is decreased. Therefore, in the case of graph (I), when the edge of the FBZ is reached, the cavity is further tuned into resonance, and the lattice thus deepens, in accordance with Fig.~\ref{fig:Fig.2}(c). In contrast, for graph (II), at the FBZ boundary the cavity is tuned away from resonance due to the increase of the coupling and hence the lattice becomes more shallow. This explains why the two oscillations are phase-shifted with respect to each other by a value close to $180^{\circ}$. The same mechanism also explains the observed decrease for trace (I) and increase for trace (II) of the mean lattice depth, which is mostly attributed to particle loss, which yields a decrease of the atom-cavity coupling $N_a \Delta_{-}$ as  time proceeds. 

\begin{figure}
\includegraphics[scale=0.45, angle=0, origin=c]{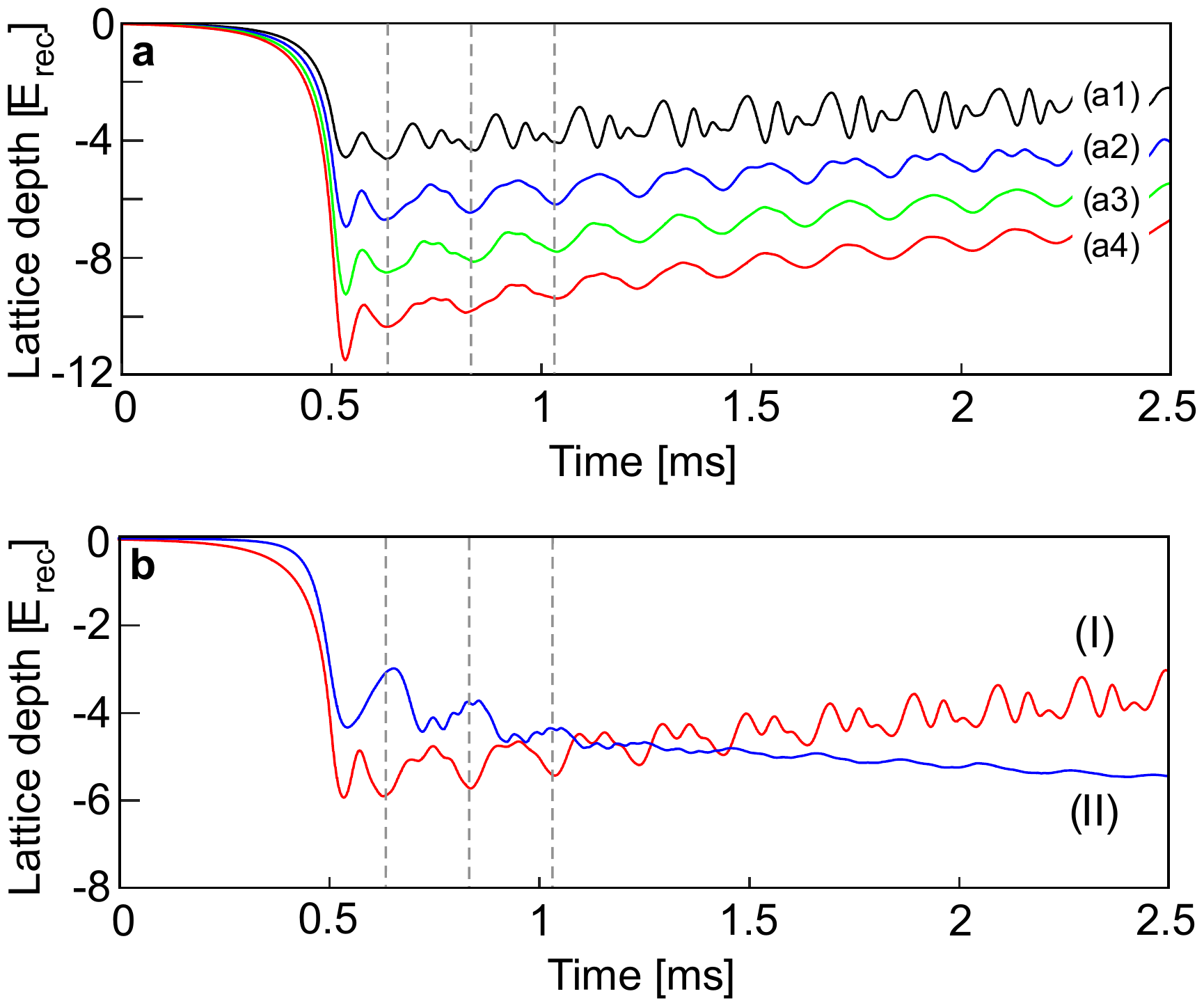}
\caption{(a) The intra-cavity light intensity is plotted in units of the induced lattice depth versus time in units of the Bloch time $T_B \equiv 2\pi / \omega_B$. The detuning is $\delta_{c} = -5.8 \, \kappa$ and $U_0$ is $-4\,E_{\mathrm{rec}}$ in (a1), $-6\,E_{\mathrm{rec}}$ in (a2), $-8\,E_{\mathrm{rec}}$ in (a3), and $-10\,E_{\mathrm{rec}}$ in (a4). (b) Calculations for opposite detunings with respect to the effective cavity resonance. The detunings with regard to the empty cavity are $\delta_{c} = -5.8 \, \kappa$ and $\delta_{c} = -1.7\,\kappa$ for trace (I) and trace (II), respectively. The values of $U_0$ were adjusted to $-6\,E_{\mathrm{rec}}$ in (I) and $-4\,E_{\mathrm{rec}}$ in (II). In (b) exponential atom loss with a rate $\Gamma \approx 100\,$s$^{-1}$ is accounted for. For all graphs $N_a = 5 \times 10^4$.}
\label{fig:Fig.3}
\end{figure}

\textbf{Model} We have set up a model based upon a Gross-Pitaevskii equation (GPE) for the matter wave function including the harmonic trap. A one-dimensional effective GPE is derived by assuming a Thomas-Fermi solution for the degrees of freedom perpendicular to the cavity axis. A time-dependent intra-cavity lattice potential is included, and the force and the harmonic confinement by the external trap along the cavity axis are implemented as time-dependent functions in accordance with the experimentally implemented protocol. The light field dynamics is modeled by a Maxwell-Bloch equation including a term that accounts for the dispersive shift of the cavity resonance caused by the atoms and a term describing the external pump beam. Details of this model are deferred to the appendix. In order to keep it tractable our model is oversimplified, i.e., the radial degrees of freedom are kept static, such that the possibility of a dynamical radial bunching of the atoms is not accounted for. Thus, we do not expect a quantitative description of our experiment, however, nevertheless the main features in our observations should be qualitatively reproduced. It turns out that the model reproduces our findings best, if we set values of the particle number and the effective detunings slightly below those believed to be adjusted experimentally. For comparable or higher values, the simulation can produce dynamic instabilities that we do not see in the experiments. We attribute this to the fact that our model cannot account for the transverse dynamics after switching off the trap in step 2 of the experimental protocol, which reduces the effective interaction strength. 

The model shows that in fact the Bloch oscillation dynamics yields an oscillation signature in the intra-cavity light field with phase offsets reproducing the detuning dependence observed in Fig.~\ref{fig:Fig.2}(c). In Fig.~\ref{fig:Fig.3}(a) the intra-cavity light intensity in units of the induced lattice depth, calculated for $\delta_{c} = -5.8 \, \kappa$, is plotted versus time. The four traces shown correspond to setting the values of the initial intra-cavity lattice depths $U_0$ (cf. appendix for definition) to be $-4\,E_{\mathrm{rec}}$ in (a1), $-6\,E_{\mathrm{rec}}$ in (a2), $-8\,E_{\mathrm{rec}}$ in (a3), and $-10\,E_{\mathrm{rec}}$ in (a4).  Note that in agreement with the upper (black) traces in Fig.~\ref{fig:Fig.2}(a) and (b), trace (a1) shows significant higher harmonics due to the shallow intra-cavity lattice. For increasing lattice depths the higher harmonic contributions decrease. In Fig.~\ref{fig:Fig.3}(b) we compare calculations of Bloch oscillation signatures in the intra-cavity light for opposite detunings with respect to the effective cavity resonance. The detunings with regard to the empty cavity are $\delta_{c} = -5.8 \, \kappa$ and $\delta_{c} = -1.7\,\kappa$ for trace (I) and trace (II), respectively. The two traces show similar qualitative behaviour as that seen in traces (I) and (II) of Fig.~\ref{fig:Fig.2}(c). As in Fig.~\ref{fig:Fig.2}(c) the oscillations in Fig.~\ref{fig:Fig.3} are phase-shifted with respect to each other by a value close to $180^{\circ}$. Note that the mean values of the lattice depths show a similar temporal behavior as in the observations in Fig.~\ref{fig:Fig.2}(c), which results from accounting for exponential atom loss $N_a(t) = N_{a,0}\,e^{-\Gamma \, t}$ with a rate $\Gamma \approx 100\,$s$^{-1}$.

Our model also provides us with momentum spectra for variable Bloch oscillation times. These are calculated after switching off the pump instantaneously and waiting during $100\,\mu$s until the exponential decay of the intra-cavity lattice taking a few cavity life times $\tau_c$ is completed. The momentum spectrum associated with trace (a2) in Fig.~\ref{fig:Fig.3}(a) is plotted into Fig.~\ref{fig:Fig.1}, where it is compared to the measured momenta given by the green triangles.

In summary, we have demonstrated a non-destructive method to monitor Bloch oscillations by using a high finesse optical cavity. Since collisions play a significant role in our experiments, we cannot observe many Bloch cycles before decoherence and losses set in. However, in connection with Feshbach tuning of the interaction to small values \cite{Koh:06, Gus:08} or the use of atoms with a naturally small scattering length like strontium \cite{Fer:06}, our method should present a major progress for efficient data sampling in experiments aiming to use Bloch oscillations as a precision force sensor, for example, to map out the gravitational acceleration.

\textbf{Appendix} 
In this section we motivate the set of mean-field equations, numerically solved in order to produce the graphs shown in Fig.~\ref{fig:Fig.1}(b) and Fig.~\ref{fig:Fig.3}. We start with the three-dimensional (3D) GPE for a BEC in an anisotropic harmonic trap. We integrate the Thomas-Fermi solution of the associated stationary GPE $\phi(\mathbf{r})$ (with the chemical potential $\mu$ and the Thomas-Fermi radii $R_{\nu}, \nu \in \{x,y,z\}$) along the cavity axis ($z$-axis) to obtain $\psi_{xy}(x,y) \equiv (\int{dz |\phi(\mathbf{r})|^2})^{1/2}$ and introduce the ansatz $\psi(\mathbf{r},t) = \psi(z,t)\, \psi_{xy}(x,y)$ into the time-dependent 3D GPE. Hence, we obtain the 1D GPE \cite{Bao:03}
\begin{align}
i \hbar \frac{d \psi}{dt} &= \left(-\frac{\hbar^2}{2m} \frac{\partial^2}{\partial z^2} + \frac{1}{2} m \omega_z^2 z^2 + g_{1d} \vert \psi \vert^2 \right) \psi 
\\ \nonumber
g_{1d} &= g_{3d} \int dx dy \vert \psi_{xy}(x,y) \vert^4, \,\, g_{3d} =\frac{4\pi N_a \hbar^2 a}{m} \,\, , 
\end{align}
where the 2D integral is carried out within the ellipse in the $xy$-plane given by the radii $R_x$ and $R_y$ and $a$ denotes the $s$-wave scattering length. Inserting the explicit form of $\phi(\mathbf{r})$ leads to $g_{1d} = \frac{4 \pi \mu^2}{9 g_{3d}}R_z^2 R_x R_y$. We may extend this equation to account for the cavity and the external force and add the Maxwell-Bloch equation for the intra-cavity field, obtaining
\begin{eqnarray}
\label{eq:matter}
i \hbar \frac{d \psi}{dt} = \Big[-\frac{\hbar^2}{2m} \frac{\partial^2}{\partial z^2} + \hbar \Delta_{-} \vert \alpha(t) \vert^2 \cos^2(k z) \quad \quad \quad \quad  
\\ \nonumber
 + \, \frac{1}{2} m \omega_z(t)^2 z^2 + F(t)\,z + g_{1d} \vert \psi(z,t) \vert^2  \Big] \psi(z,t)
\\ \label{eq:light}
 \frac{d \alpha}{dt} =  \left(-\kappa + i \delta_c(t) -i N_a \Delta_{-} \langle \cos^2(k z) \rangle_{\psi} \right) \alpha (t) + \eta(t)
 \\ \nonumber
 \langle \cos^2(k z) \rangle_{\psi} \equiv \int dz \ \vert \psi(z,t) \cos(k z) \vert^2\,\, .
 \end{eqnarray}
The time-dependences of $F(t)$, $\omega_z(t)$, $\eta(t)$ and $\delta_c(t)$ are chosen according to the experimental protocol. By means of the steady-state solution of eq.~\ref{eq:light}, we can evaluate the intra-cavity lattice depth $U_0(\eta, \delta_c)$ introduced by a pump field with strength $\eta$ and detuning $\delta_c$, assuming that the BEC shape is that expected in the initial harmonic trap, yielding
\begin{eqnarray}
\label{eq:eta}
U_0 = \frac{\eta^2 \, \Delta_{-} }{ \kappa^2 + \left(\delta_c - N_a \Delta_{-} \langle \cos^2(k z) \rangle_{\psi(z,t)} \right)^2}\,\,.
 \end{eqnarray}
This lets us characterize our pump beam adjustment via specifying the more physical parameter $U_0$ instead of $\eta$. 

\textbf{Acknowledgments.}
This work was partially supported by DFG-SFB925 and the Hamburg centre of ultrafast imaging (CUI).

\end{document}